\newcommand{\CLASSINPUTtoptextmargin}{0.75in}%
\newcommand{\CLASSINPUTbottomtextmargin}{1in}%
\newcommand{\CLASSINPUTinnersidemargin}{0.63in}%
\newcommand{\CLASSINPUToutersidemargin}{0.63in}%
\documentclass[conference,10pt,letterpaper]{IEEEtran}% 
\usepackage{amsmath}% for double integral symbol in this template
\usepackage{graphicx}% for figures
\usepackage{multirow}% to allow multiple-row elements in tabular environment
\usepackage[none]{hyphenat}% turn off hyphenation to make text extraction and indexing easier
\usepackage{float}% better control of floating figures and tables
\usepackage{subfig}% for subfigures within figures
\usepackage{dblfloatfix}% fix to allow page-wide floats at bottom of page
\usepackage{algorithm}
\usepackage{algpseudocode}
\usepackage{booktabs} 
\usepackage{xcolor}
\usepackage{verbatim}
\usepackage{url}
\usepackage{hyperref}

\makeatletter
\renewcommand{\ALG@beginalgorithmic}{\normalsize} % Ensures normal size, but no numbers
\algrenewcommand\alglinenumber[1]{} % Disables line numbers explicitly
\makeatother

\usepackage{t1enc}% allows access to various special characters
\usepackage{times}% change font to Nimbus Roman (based on Times Roman)
\makeatletter

\def\@IMSauthorblockNAMEstyle{\normalfont\IMSauthorsize}
\def\@IMSauthorblockAFFILstyle{\normalfont\IMSaffilsize}
\def\@IMSauthorblockEMAILstyle{\normalfont\IMSaffilsize}
\def\IMSauthorblockNAME#1{%
\relax\@IMSauthorblockNAMEstyle%
#1%
}%
\def\IMSauthorblockAFFIL#1{%
\relax\@IMSauthorblockAFFILstyle%
\vskip\@IEEEauthorblockAtopspace
#1%
}%
\def\IMSauthorblockEMAIL#1{%
\relax\@IMSauthorblockEMAILstyle%
\vskip\@IEEEauthorblockAtopspace
#1%
}%
\newcommand{\IMSauthor}[1]{%
\ifIsBlindReviewVersion%
\author{\phantom{\parbox{\textwidth}{\center\relax#1}}}%
\else%
\author{\parbox{\textwidth}{\center\relax#1}}%
\fi%
}%
\newif\ifIsBlindReviewVersion
\def\IMSthispaperforblindreview{\IsBlindReviewVersiontrue}
\def\IMSthispaperforfinalpublication{\IsBlindReviewVersionfalse}
\def\@maketitle{\newpage
\bgroup\par\addvspace{0.5\baselineskip}\centering%
\ifCLASSOPTIONtechnote% technotes
   {\bfseries\large\@IEEEcompsoconly{\sffamily}\@title\par}\vskip 1.3em{\lineskip .5em\@IEEEcompsoconly{\sffamily}\@author
   \@IEEEspecialpapernotice\par{\@IEEEcompsoconly{\vskip 1.5em\relax
   \@IEEEtitleabstractindextextbox{\@IEEEtitleabstractindextext}\par
   \hfill\@IEEEcompsocdiamondline\hfill\hbox{}\par}}}\relax
\else% not a technote
   \vskip0.2em{\IMStitlesize\ifCLASSOPTIONtransmag\bfseries\LARGE\fi\@IEEEcompsoconly{\sffamily}\@IEEEcompsocconfonly{\normalfont\normalsize\vskip 2\@IEEEnormalsizeunitybaselineskip
   \bfseries\Large}\@title\par}\vskip1.0em\par% CAUSAL PRODUCTIONS change on this line
   \ifCLASSOPTIONconference%
      {\@IEEEspecialpapernotice\mbox{}\vskip\@IEEEauthorblockconfadjspace%
       \mbox{}\hfill\begin{@IEEEauthorhalign}\@author\end{@IEEEauthorhalign}\hfill\mbox{}\par}\relax
   \else% peerreviewca, peerreview or journal
      \ifCLASSOPTIONpeerreviewca
         {\@IEEEcompsoconly{\sffamily}\@IEEEspecialpapernotice\mbox{}\vskip\@IEEEauthorblockconfadjspace%
          \mbox{}\hfill\begin{@IEEEauthorhalign}\@author\end{@IEEEauthorhalign}\hfill\mbox{}\par
          {\@IEEEcompsoconly{\vskip 1.5em\relax
           \@IEEEtitleabstractindextextbox{\@IEEEtitleabstractindextext}\par\hfill
           \@IEEEcompsocdiamondline\hfill\hbox{}\par}}}\relax
      \else% journal, peerreview or transmag
         \ifCLASSOPTIONtransmag
           {\@IEEEspecialpapernotice\mbox{}\vskip\@IEEEauthorblockconfadjspace%
            \mbox{}\hfill\begin{@IEEEauthorhalign}\@author\end{@IEEEauthorhalign}\hfill\mbox{}\par
           {\vspace{0.5\baselineskip}\relax\@IEEEtitleabstractindextextbox{\@IEEEtitleabstractindextext}\vspace{-1\baselineskip}\par}}\relax
         \else% journal or peerreview
           {\lineskip.5em\@IEEEcompsoconly{\sffamily}\sublargesize\@author\@IEEEspecialpapernotice\par
           {\@IEEEcompsoconly{\vskip 1.5em\relax
            \@IEEEtitleabstractindextextbox{\@IEEEtitleabstractindextext}\par\hfill
            \@IEEEcompsocdiamondline\hfill\hbox{}\par}}}\relax
         \fi
      \fi
   \fi
\fi\par\addvspace{0.0\baselineskip}\egroup}% CAUSAL PRODUCTIONS change on this line, reduce the vspace from 0.5\baselineskip to 0.0

\def\IMStitlesize{\@setfontsize{\IMStitlesize}{18}{21pt}}% CAUSAL PRODUCTIONS change on this line
\def\IMSauthorsize{\@setfontsize{\IMSauthorsize}{12}{13pt}}% CAUSAL PRODUCTIONS change on this line
\def\IMSaffilsize{\@setfontsize{\IMSaffilsize}{12}{13pt}}% CAUSAL PRODUCTIONS change on this line
\def\IMScaptionsize{\@setfontsize{\IMScaptionsize}{8}{9pt}}% CAUSAL PRODUCTIONS change on this line
\def\IMSbibsize{\@setfontsize{\IMSbibsize}{8}{9pt}}% CAUSAL PRODUCTIONS change on this line

\def\@IEEEauthorblockNstyle{\IMSauthorsize\@IEEEcompsocnotconfonly{\sffamily}\@IEEEcompsocconfonly{\large}}%CAUSAL PRODUCTIONS removed sublargesize to get correct IMSauthorsize
\def\@IEEEauthorblockAstyle{\IMSaffilsize\@IEEEcompsocnotconfonly{\sffamily}\@IEEEcompsocconfonly{\itshape}\@IEEEcompsocconfonly{\large}}%CAUSAL PRODUCTIONS removed normalsize to get correct IMSaffilsize
\def\@IEEEauthordefaulttextstyle{\IMSauthorsize\@IEEEcompsocnotconfonly{\sffamily}\sublargesize}%CAUSAL PRODUCTIONS

\def\thebibliography#1{\section*{\refname}%
    \addcontentsline{toc}{section}{\refname}%
    \IMSbibsize\@IEEEcompsocconfonly{\small}\vskip 0.3\baselineskip plus 0.1\baselineskip minus 0.1\baselineskip% CAUSAL PRODUCTIONS change on this line
    \list{\@biblabel{\@arabic\c@enumiv}}%
    {\settowidth\labelwidth{\@biblabel{#1}}%
    \leftmargin\labelwidth
    \advance\leftmargin\labelsep\relax
    \itemsep \IEEEbibitemsep\relax
    \usecounter{enumiv}%
    \let\p@enumiv\@empty
    \renewcommand\theenumiv{\@arabic\c@enumiv}}%
    \let\@IEEElatexbibitem\bibitem%
    \def\bibitem{\@IEEEbibitemprefix\@IEEElatexbibitem}%
\def\newblock{\hskip .11em plus .33em minus .07em}%
\ifCLASSOPTIONtechnote\sloppy\clubpenalty4000\widowpenalty4000\interlinepenalty100%
\else\sloppy\clubpenalty4000\widowpenalty4000\interlinepenalty500\fi%
    \sfcode`\.=1000\relax}

\long\def\@makecaption#1#2{%
\ifx\@captype\@IEEEtablestring%
\par\@IEEEtabletopskipstrut% strut used to align table caption with facing column
\else
\@IEEEfigurecaptionsepspace
\fi
\setbox\@tempboxa\hbox{\normalfont\IMScaptionsize {#1.}\nobreakspace\nobreakspace #2}%
\ifdim \wd\@tempboxa >\hsize%
\setbox\@tempboxa\hbox{\normalfont\IMScaptionsize {#1.}\nobreakspace\nobreakspace}%
\parbox[t]{\hsize}{\normalfont\IMScaptionsize\noindent\unhbox\@tempboxa#2}%
\else
\ifCLASSOPTIONconference \hbox to\hsize{\normalfont\IMScaptionsize\hfil\box\@tempboxa\hfil}%
\else \hbox to\hsize{\normalfont\IMScaptionsize\box\@tempboxa\hfil}%
\fi\fi
\ifx\@captype\@IEEEtablestring%
\@IEEEtablecaptionsepspace
\else
\fi}

\newlength\tablecaptiontotableskip
\newlength\figuretocaptionskip
\def\@IEEEfigurecaptionsepspace{\vskip\figuretocaptionskip\relax}%
\def\@IEEEtablecaptionsepspace{\vskip\tablecaptiontotableskip\relax}%

\def\abstract{\normalfont%
\@IEEEabskeysecsize\bfseries\textit{\abstractname}\,\bfseries\textit{---}\,%
\@IEEEgobbleleadPARNLSP}%

\def\IEEEkeywords{\normalfont%
\@IEEEabskeysecsize\bfseries\textit{\IEEEkeywordsname}\,\bfseries\textit{---}\,%
\@IEEEgobbleleadPARNLSP}%
\def\endIEEEkeywords{\relax\vspace{0.67ex}%
\par\if@twocolumn\else\endquotation\fi%
\normalsize\normalfont}%

\DeclareRobustCommand*{\IMSauthorrefmark}[1]{\raisebox{0pt}[0pt][0pt]{\textsuperscript{\footnotesize{#1}}}}%
\def\@IEEEauthorblockNtopspace{0ex}
\def\@IEEEauthorblockAtopspace{1mm}
\def\tablename{Table}% IMS: was TABLE in IEEETran, but we are now using capitalized caption text not smallcaps
\def\thetable{\arabic{table}}% IMS: Tables are numbered using arabic not roman
\def\IEEEkeywordsname{Keywords}% use Keywords instead of Index Terms
\def\subsubsection{\@startsection{subsubsection}{3}{\z@}{1.5ex plus 1.5ex minus 0.5ex}%
{0.7ex plus .5ex minus 0ex}{\normalfont\normalsize\itshape}}%
\def\@seccntformat#1{\csname the#1dis\endcsname\relax}% moved the spacer \hskip 0.5em to individual handlers below
\def\thesectiondis{\thesection.\hskip 0.5em}%					I.	% CAUSAL PRODUCTIONS: moved the hskip spacer from \@seccntformat to here
\def\thesubsectiondis{{\hbox to\parindent{\Alph{subsection}.}}}%		B.	% CAUSAL PRODUCTIONS: indent the subsection name to match paragraph indent
\def\thesubsubsectiondis{{\hbox to \parindent{\arabic{subsubsection})}}}%	3)	% CAUSAL PRODUCTIONS: indent the subsubsection name to match paragraph indent
\def\theparagraphdis{{\hbox to \parindent{\alph{paragraph})}}}%			d)	% CAUSAL PRODUCTIONS: indent the subsubsubsection name to match paragraph indent
\IEEEilabelindentA \parindent
\IEEEilabelindent \IEEEilabelindentA
\IEEEelabelindent \parindent
\IEEEdlabelindent \parindent
\IEEElabelindent \parindent
\newlength\@IMSparindent
\newcommand\IMSdisplayacksection[1]{%
\ifIsBlindReviewVersion%
\noindent\phantom{\parbox[t]{\columnwidth}{\normalbaselines\setlength{\parindent}{\@IMSparindent}{#1}\strut}}%\IMSacktext
\else%
\noindent\parbox[t]{\columnwidth}{\normalbaselines\setlength{\parindent}{\@IMSparindent}{#1}\strut}%
\fi%
}%

\makeatother

\usepackage{cite}
\begin{document}
%%%%%%%%%%%%%%%%%%%%%%%%%%%%%%%%%%%%%%%%%%%%%%%%%%%%%%%%%%%%%%%%%%%%%%%%%%%%%
% We use \raggedbottom to avoid latex adding vertical space around headings.
% This gives a better idea to the author about how much white space remains
% as the page limit is approached.
\raggedbottom
%
%%%%%%%%%%%%%%%%%%%%%%%%%%%%%%%%%%%%%%%%%%%%%%%%%%%%%%%%%%%%%%%%%%%%%%%%%%%%%
% PAPER TITLE AND AUTHOR BLOCK
%
% The paper title can use linebreaks \\ within to get better formatting if desired.
%
%\title{Ultra-Fast/Fine Inverse Design for High-Performance Electromagnetic Applications: Filter/Diplexer and Complexer Impedance Power Combiners}

\title{A Single-Tuning-Element Loaded Pixelated Tunable Band-Pass Filters with Low Loss via Inverse Synthesis on Massive-Scale Dataset}
%
% Next we define the author names and affiliations.
% Author names are listed using \IMSauthorblockNAME{} with comma separators between names.
% Affiliations arce listed using \IMSauthorblockAFFIL{} with \\ separators between affiliations.
% Email addresses are listed using \IMSauthorblockEMAIL{} with comma separators between emails.
% See below for examples of each of these.
%
% Symbols marking author-affiliation relations are output using \IMSauthorrefmark{}.
%
% Next we typeset the authorblock either as visible text, or as an empty
% box of the same size, based on the value of the Blind Review Flag.
% Note that the Blind Review Flag also determines whether the Acknowledgments
% section is visible or invisible.
% To set the flag to Blind Review mode, simply uncomment the next line
\IMSthispaperforblindreview
% or to set the flag to Final Paper mode (with author block visible) then
% simply uncomment the next line:
\IMSthispaperforfinalpublication
\IMSauthor{%
\IMSauthorblockNAME{% Author Names
Woojun Lee\IMSauthorrefmark{\#1},
Pouya Faeghi\IMSauthorrefmark{\#2}, and
Jeffrey Sean Walling\IMSauthorrefmark{\#3}
}% end of \IMSauthorblockNAME
\\%
\IMSauthorblockAFFIL{% Author Affiliations
\IMSauthorrefmark{\#}Virginia Tech
}% end of \IMSauthorblockAFFIL
\\%
\IMSauthorblockEMAIL{% Author Emails
\{\IMSauthorrefmark{1}woojun, \IMSauthorrefmark{2}pfaeghi, \IMSauthorrefmark{3}jswalling\}@vt.edu
}% end of \IMSauthorblockEMAIL
}% end of \IMSauthor
%
% Next we make the title/author block using the information defined above.
\maketitle
%
%%%%%%%%%%%%%%%%%%%%%%%%%%%%%%%%%%%%%%%%%%%%%%%%%%%%%%%%%%%%%%%%%%%%%%%%%%%%%
% ABSTRACT paragraph.
%
% As a general rule, do not put math, special symbols or citations
% in the abstract paragraph.
%
\begin{abstract}
This paper presents a novel three-phase inverse-design flow for synthesizing tunable pixelated band-pass filters. The workflow consists of: 1) acquisition of a massive dataset (\(\sim\!300{,}000\) 5-port S-parameter samples) using the group’s custom electromagnetic solvers accelerated by method-of-moments pre-computation; 2) identification of an initial seed through brute-force search with dataset augmentation by dynamically reassigning the input, output, and tuning ports and randomly applying open or short terminations to the remaining ports; and 3) fine-tuning of the selected seed using a direct-binary-search algorithm. Two design targets are considered: 4G mid-/high-band tunable band-pass filters and 5G n79/UNIII-band tunable band-pass filters. One designed filter achieves a \(28\%\) tuning range, with passband peaks reconfigured from \(4.64\) to \(6.16\)~GHz and insertion loss from \(1.3\) to \(1.8\)~dB in simulation. All inversely designed filters use only a single varactor, representing the first reported demonstration of continuously reconfigurable inverse design in the microwave regime.

\end{abstract}
%
%%%%%%%%%%%%%%%%%%%%%%%%%%%%%%%%%%%%%%%%%%%%%%%%%%%%%%%%%%%%%%%%%%%%%%%%%%%%%
% THE REST OF THE PAPER follows.
%

\section{Introduction}

Tunable bandpass filters (BPFs) are getting attractive since multiple wireless standards ranging from 3G/4G/5G and WIFI 5/6/7 are emerging. Tunable BPFs are to decrease the number of filters. One of the popular techniques to design tunable filters is loading multiple tuning elements such as varactors or p-i-n diodes to microstrip or substrate integrated wavguide structures \cite{yang2015tunable, chen2018microstrip}. Most of the designs contain 4-10 elements, which limits insertion loss, while some designs loaded with as low as two elements with moderate tuning ranges are reported in \cite{two_1}.

Meanwhile, microwave community is witnessing an emerging paradigm, inverse design, of passive electromagnetic components such as BPFs, matching networks, and transformers, which mostly take the form of pixelated traces on a ground plane \cite{jungmin, jungmin_APMC, prince}. While most of the research is conducted to increase bandwidth of the power amplifier, compact BPFs are reported in \cite{hongkong, jungmin}. 

To date, inverse-design techniques have not been fully extended to reconfigurable passive devices such as tunable BPFs. This direction is important because the vast design space formed by active components and complex electromagnetic structures remains largely unexplored, whereas most tunable microwave devices still rely on conventional transmission-line and coupled-resonator or coupler-based topologies. In addition, the complex interconnects and coupling among pixelated traces may enable high-performance tunable BPFs with only a small number of tuning elements.

In this work, a novel three-step workflow is proposed for synthesizing tunable BPFs with a single varactor using a massive dataset of \(\sim\!300{,}000\) unloaded 5-port S-parameter samples. The goal is to demonstrate that highly complex pixelated traces combined with active components can realize reconfigurable passband and stopband responses. The synthesized BPFs will be shown to achieve a moderate tuning range with lower loss than state-of-the-art counterparts using a larger number of tuning elements, despite relying on only a single tuning element. In contrast to most previous inverse-design approaches, such as \cite{prince,jungmin_APMC}, which combine neural-network surrogate models with search algorithms, the proposed method directly searches for candidate designs from the dataset. This workflow is made possible by the large dataset size, which is sufficient to contain promising initial seeds. To the best of the authors' knowledge, this is the first report of continuously reconfigurable inverse design in the microwave regime.

%%%%%%%%%%%%%%%%%%%%%%%%%%%%%%%%%%%%%%%%%%%%%%%%%%%%%%%%%%%%%%%%%%%%%%%%%%%%%

\section{Inverse Synthesis Based on Massive Data-set}
%It is an extremely challenging task to find a highly arbitrary surfaces that yield a tunable filtering response from scratch. Therefore, a novel systematic three-step inverse-synthesis methodology, leveraging a massive dataset to enable the design of reconfigurable passive devices is detailed in this section. 

A procedure to find a highly arbitrary surfaces that yield a tunable filtering response from scratch consists of three stages: 1) data acquisition, in which the dataset size is augmented through randomized port assignment and short/open loading; 2) selection of an initial seed exhibiting widely distributed band-pass response peaks; and 3) extensive optimization of the chosen seed through the tree-search algorithm. Overall work flow is outlined in Fig. 1. 

\begin{figure*}[t]
    \centering
    \includegraphics[width=1\textwidth]{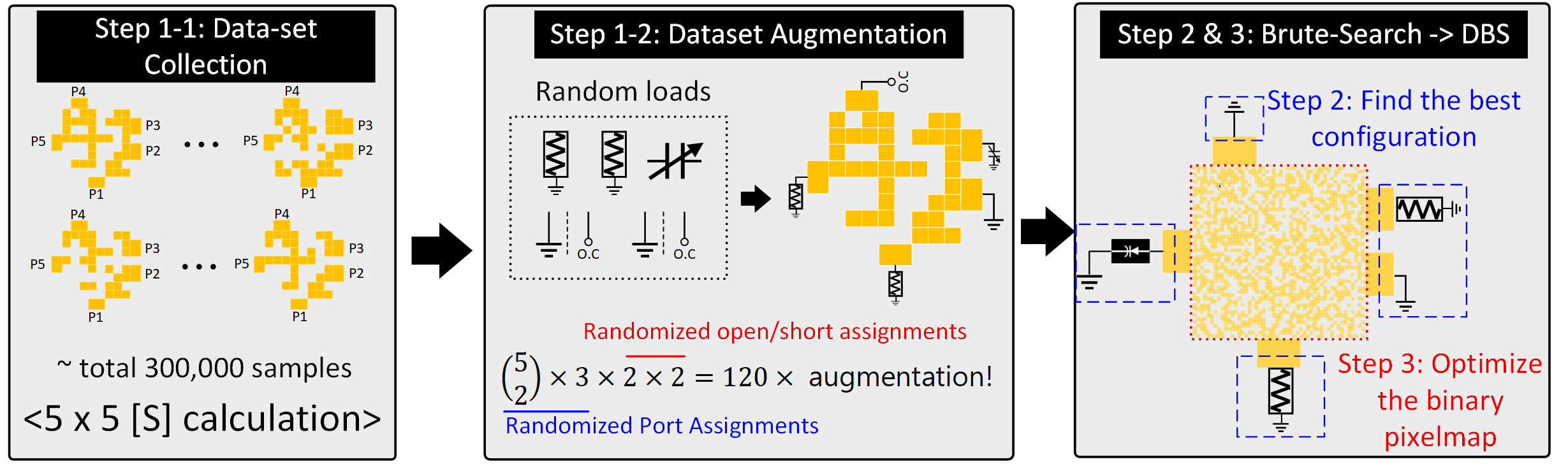}
    \caption{Three-step workflow to inversely synthesize tunable BPFs. Step 1 involves the generation of massive data-set and augmentation with a factor of 120. Follwing steps include the brute search of the randomized loadmaps and pixelmaps to find an initial seed and a direct binary search algorithm to fine-tune the pixelmap.}
\label{fig::fig1}
\end{figure*}

%%%%%%%%%%%%%%%%%%%%%%%%%%%%%%%%%%%%%%%%%%%%%%%%%%%%%%%%%%%%%%%%%%%%%%%%%%%%%
\subsection{Data-set Acquistion with Augmentation}
A representative geometry used for dataset generation is shown in Fig.~2. Random $50 \times 50$ pixel maps with a pixel size of $0.2$~mm and five excitation ports are sampled with fill densities ranging from $0.4$ to $0.8$. As will be shown later, densities between $0.6$ and $0.7$ are particularly well suited for identifying structures that exhibit a single passband. The frequency range extends from DC to $10$~GHz. To accelerate dataset generation, the pre-computation technique reported in~\cite{gamom, woojun_arxiv} is employed. With eight NVIDIA L40S GPUs, a 5-port S-parameter response with 150 frequency points over DC--10~GHz can be generated in approximately $0.5$~seconds.

The unloaded 5-port S-parameters are then augmented by randomly assigning the input/output ports and varactor ports, resulting in a $30\times$ increase in the search space. The dataset is further augmented by randomly applying either an open or a short termination to the remaining two ports, leading to an overall $120\times$ increase in the search space. A similar strategy was used in~\cite{prince} to derive six 2-port S-parameter responses from a single 4-port network. 

A concise mathematical procedure for constructing the loaded 3-port scattering matrix \([S]_{\mathrm{loaded},3\mathrm{p}}\) from the unloaded 5-port scattering matrix \([S]_{\mathrm{unloaded},5\mathrm{p}}\) is described as follows. Assume that Ports~1 and 2 denote the selected input and output ports, Port~3 denotes the varactor port, and Ports~4 and 5 denote the remaining auxiliary ports, which are terminated by either an open or a short. To obtain the unloaded 3-port network, \([S]_{\mathrm{unloaded},5\mathrm{p}}\) is first transformed into either the impedance matrix \([Z]\) or the admittance matrix \([Y]\). All of the network parameters follow the definition in \cite{pozar2011microwave}. 

When a port is terminated with an ideal open, it can be eliminated in the impedance domain by deleting the corresponding row and column of \([Z]\). Similarly, when a port is terminated with an ideal short, it can be eliminated in the admittance domain by deleting the corresponding row and column of \([Y]\). The reduced matrix is then converted back to the scattering domain and denoted by \([S]_{\mathrm{unloaded},3\mathrm{p}}\). For example, if one auxiliary port is open-terminated and the other is short-terminated, \([S]_{\mathrm{unloaded},5\mathrm{p}}\) is first converted to \([Z]_{5\mathrm{p}}\), from which the open-terminated port is removed to obtain \([Z]_{4\mathrm{p}}\). The reduced matrix is then converted to \([Y]_{4\mathrm{p}}\), and the short-terminated port is removed to obtain \([Y]_{3\mathrm{p}}\).

Afterward, the varactor port is terminated by a one-port load with reflection coefficient
\begin{equation}
\Gamma_{L}(f)=\frac{Z_{L}(f)-Z_{0}}{Z_{L}(f)+Z_{0}},
\end{equation}
where \(Z_{0}=50~\Omega\), and $Z_{L}(f)$ represents the impedance of a varactor. Partitioning \(\mathbf{S}_{3}(f)\) as
\begin{equation}
\mathbf{S}_{unloaded,3p}=
\begin{bmatrix}
\mathbf{S}_{1\&2,1\&2} & \mathbf{S}_{1\&2, 3}\\
\mathbf{S}_{3, 1\&2} & \mathbf{S}_{33}
\end{bmatrix},
\end{equation}
where $\mathbf{S}_{i,j}$ is each sub-matrix of $\mathbf{S}_{unloaded,3p}$ associated with $i$-th and $j$-th entry. Then, the final loaded 2-port network is given as below \cite{port_reduction}
\begin{equation}
\mathbf{S}_{loaded,2p}
=
\mathbf{S}_{12,12}
+
\mathbf{S}_{12, 3}
\left(1-\Gamma_{L}(f)\mathbf{S}_{33}\right)^{-1}
\Gamma_{L}(f)\mathbf{S}_{3, 12}.
\end{equation}
In total, approximately $300{,}000$ unloaded 5-port S-parameter samples are generated in about $20$~hours. After augmentation, this corresponds to a dataset of approximately $36{,}000{,}000$ samples.

%%%%%%%%%%%%%%%%%%%%%%%%%%%%%%%%%%%%%%%%%%%%%%%%%%%%%%%%%%%%%%%%%%%%%%%%%%%%%
\subsection{Selection of Initial Seeds and Fine-tuning}
\begin{figure}[t!]
    \centering
    \includegraphics[width=1\columnwidth]{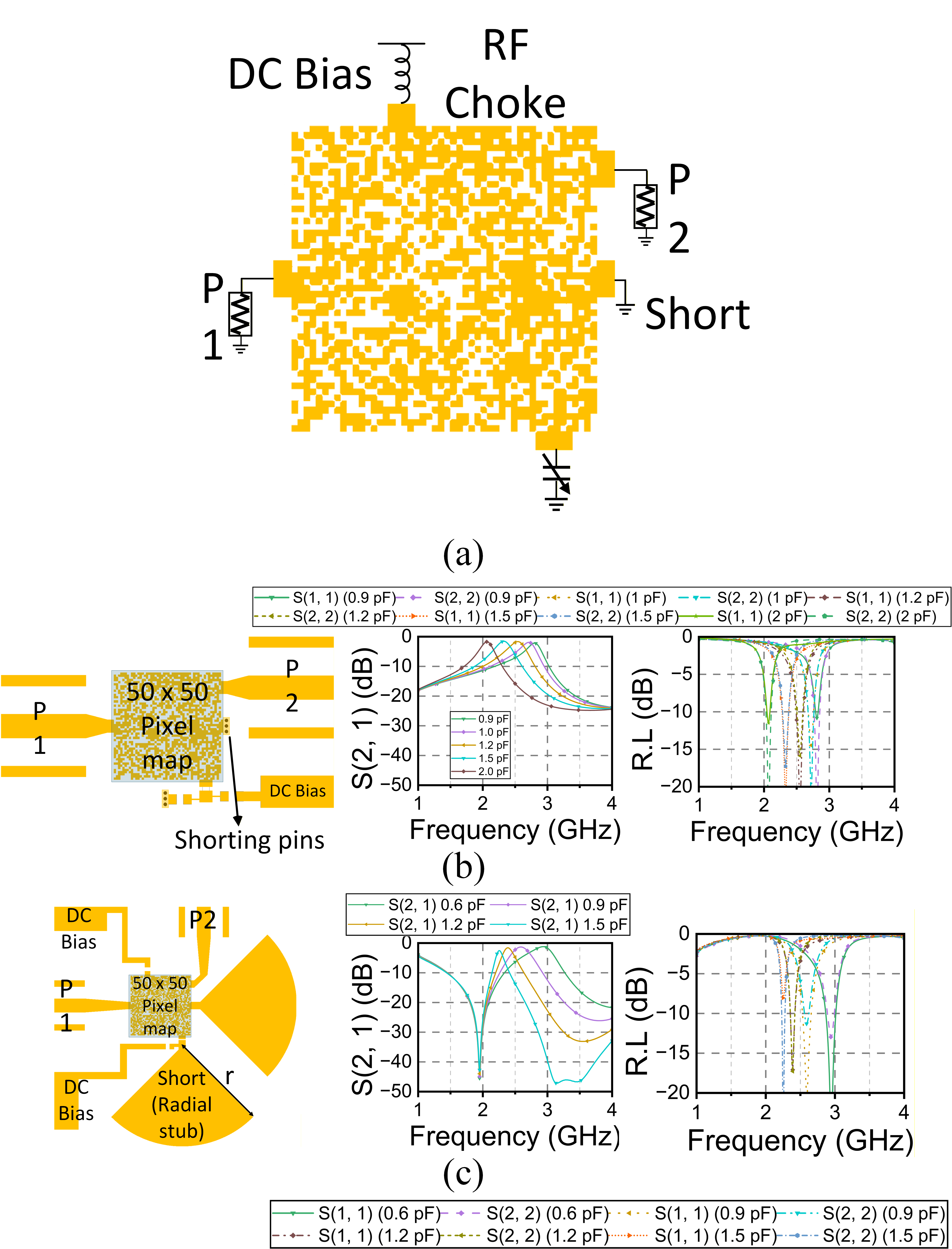}
    \caption{(a) Optimized configuration that targets 4G MB/HB band. (b) Simulated S-parameter response of the designed filters with the shorting pins at the shorted port. (c) Simulated S-parameter response of the designed filters with the radial stub at the shorted port ($r=12.1~\mathrm{mm}$).}
\label{fig::fig2}
\end{figure}

The second step is design sifting to identify a potential wide-tuning-range tunable BPFs. A mono-objective figure of merit (FoM) is employed to identify candidates with wide tuning range and low insertion loss. For each loading state, the transmission response is evaluated through \(S_{21}\) over one of two capacitance ranges: seven sampled values from \(0.16\)~pF to \(1\)~pF, or seven sampled values from \(0.44\)~pF to \(4.71\)~pF. A state is regarded as valid only if it exhibits a single passband above \(-3\) dB, the corresponding passband bandwidth is less than \(1\) GHz, and all other spurious peaks remain below \(-8\) dB. In addition, at least \(40\%\) of all loading states must satisfy these hard constraints; otherwise, the candidate is discarded.

For the valid states, the FoM is defined using only the peak \(S_{21}\) values and their corresponding frequency points as
\begin{equation}
\mathrm{FoM} = w_f \Delta f + w_p \overline{S}_{21,\mathrm{peak}},
\end{equation}
where \(\Delta f\) is the peak-frequency spread across all valid states, $\overline{S}_{21,\mathrm{peak}}$ is the average main-peak level in dB, \(w_f=1\), and \(w_p=0.35\). Maximizing this FoM favors candidates with broadly distributed passband peaks and low insertion loss. At this stage, the rejection level away from the main peak is not explicitly included in the FoM, in order to avoid excluding potentially promising wide-tuning-range candidates.

Afterward, the final step is an extensive optimization stage to fine-tune the initial seed. A direct-binary-search (DBS) algorithm~\cite{jungmin} is employed using the modified FoM
\begin{equation}
\mathrm{FoM}
=
\alpha \Delta f
+
\beta P_{\mathrm{rej}}
+
\gamma P_{\mathrm{IL}}
+
\delta P_{\mathrm{RL}},
\end{equation}
where \(\Delta f\) denotes the peak-frequency spread across all valid capacitance states. The penalty terms are defined as
\begin{equation}
P_{\mathrm{rej}}=\sum_{C} R_{\mathrm{rej}}(C),
\end{equation}
\begin{equation}
P_{\mathrm{IL}}=\sum_{C} R_{\mathrm{IL}}(C),
\end{equation}
\begin{equation}
P_{\mathrm{RL}}=\sum_{C}\left(R_{\mathrm{RL},11}(C)+R_{\mathrm{RL},22}(C)\right),
\end{equation}
with
\begin{equation}
R_{\mathrm{IL}}(C)=\left|S_{21,\mathrm{pk}}(C)\right|,
\end{equation}
\begin{equation}
R_{\mathrm{RL},11}(C)=\max\!\left(0,\,15-\left|S_{11,\mathrm{pk}}(C)\right|\right),
\end{equation}
\begin{equation}
R_{\mathrm{RL},22}(C)=\max\!\left(0,\,15-\left|S_{22,\mathrm{pk}}(C)\right|\right),
\end{equation}
and
\begin{equation}
R_{\mathrm{rej}}(C)
=
\frac{1}{N_C}
\sum_{f\in\mathcal{F}_{\mathrm{rej}}(C)}
\left|S_{21}(f,C)\right|,
\end{equation}
where \(S_{21,\mathrm{pk}}(C)\), \(S_{11,\mathrm{pk}}(C)\), and \(S_{22,\mathrm{pk}}(C)\) denote the peak values in dB at capacitance state \(C\), and \(\mathcal{F}_{\mathrm{rej}}(C)\) denotes the out-of-band frequency samples excluding the main passband and a \(\pm 0.5\)~GHz window around the peak frequency. It is noted that the center frequency is dynamically reset by identifying the peak frequency points across different states. Hard constraints are also enforced so that each capacitance state must exhibit a single passband above \(-6\)~dB with bandwidth below \(1\)~GHz. In this work, weighting factors of\(\alpha=100\), \(\beta=10\), \(\gamma=5\), and \(\delta=1\) are chosen.

%%%%%%%%%%%%%%%%%%%%%%%%%%%%%%%%%%%%%%%%%%%%%%%%%%%%%%%%%%%%%%%%%%%%%%%%%%%%%
\section{Demonstration of Pixelated Tunable BPFs}

\begin{figure}[t!]
    \centering
    \includegraphics[width=1\columnwidth]{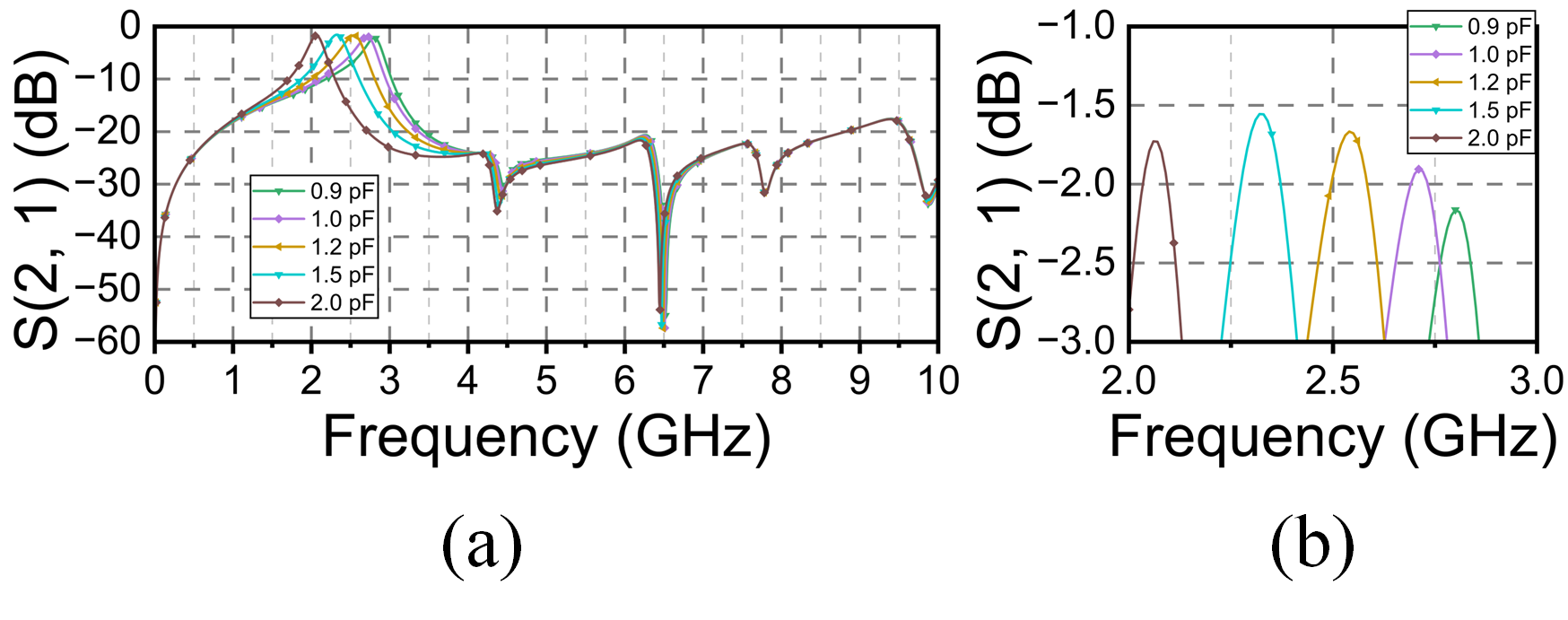}
    \caption{Wideband/narrow-band response of the filter in  Fig. 2(b).}
\label{fig::fig3}
\end{figure}

\begin{figure}[t!]
    \centering
    \includegraphics[width=1\columnwidth]{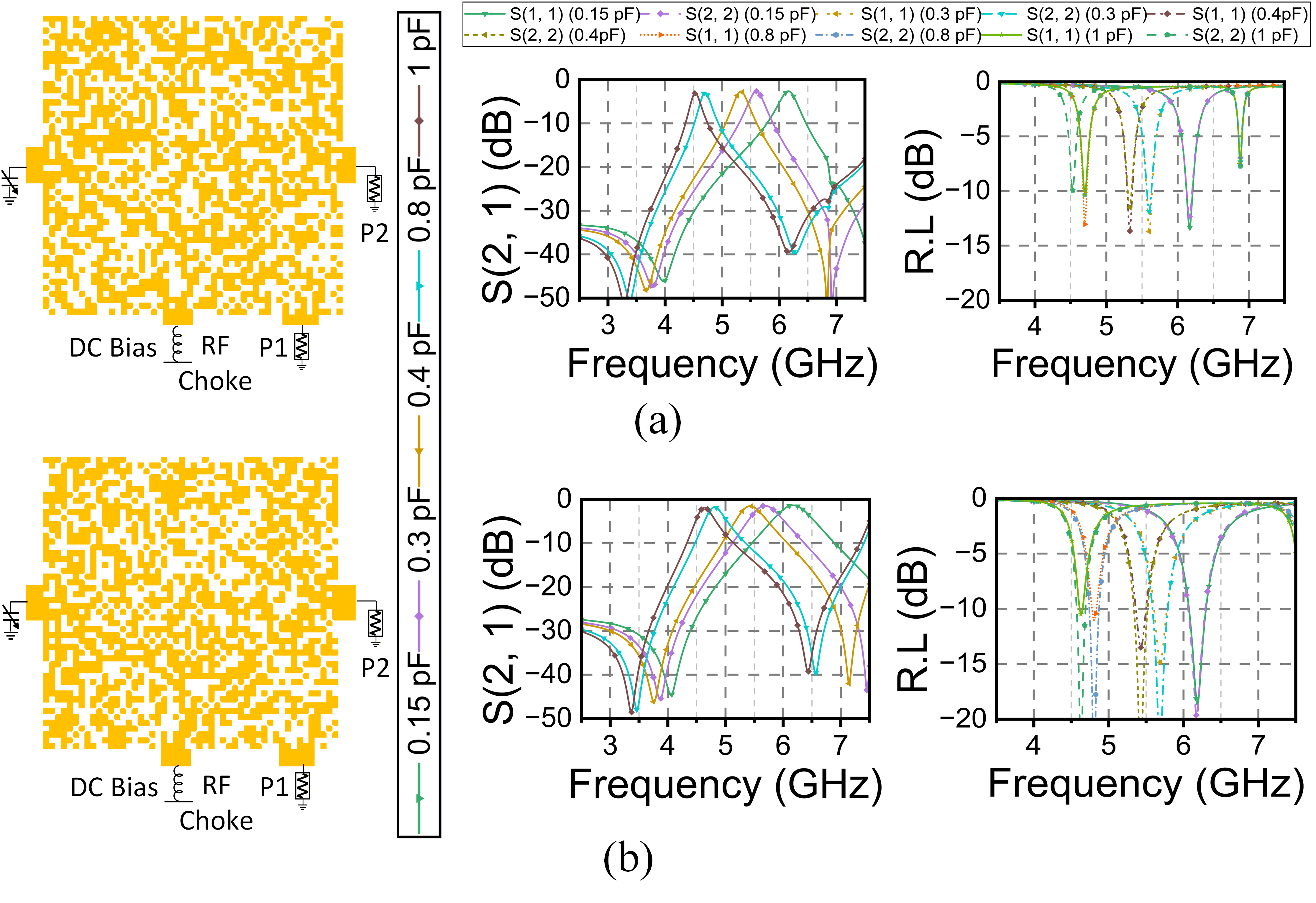}
    \caption{Simulated S-parameter response of the filters with different configuration, targeting 5G n79/UNIII band.}
\label{fig::fig4}
\end{figure}

The first target band is from 1.7 GHz to 2.69 GHz, spanning 4G mid-band (MB) and high-band (HB). The second target band is ranging from 5G n79 bands to UNII-bands. The most tunable filters focus on the former one. This work targets to tackle the challenging task of designing low-loss band-pass filters at relatively high bands, notably with a single tuning varactor.  While numerous tunable filters are reported to target this band \cite{chen2018microstrip, yang2015tunable, two_1}, no filters with a single tuning element are reported yet. 

%%%%%%%%%%%%%%%%%%%%%%%%%%%%%%%%%%%%%%%%%%%%%%%%%%%%%%%%%%%%%%%%%%%%%%%%%%%%%
%%%%%%%%%%%%%%%%%%%%%%%%%%%%%%%%%%%%%%%%%%%%%%%%%%%%%%%%%%%%%%%%%%%%%%%%%%%%%
\subsection{Layout Considerations}
In Fig 2., an exemplar layout targeting 4G MB/HB is shown. As mentioned above the pixel size is chosen as 0.2 mm, pixel map is $50\times50$ and an overall dimension is accordingly $10~\mathrm{mm} \times 10~\mathrm{mm}$. One additional shorting port other than two input/output ports is configured due to the randomized load of open and short is chosen. While one of the bias inductors should be placed nearby the varactor, the position of the varactor were placed upon the widest continuous pixel traces that have a DC connection path with the other side of the varactor. This work also studies the effect of replacing the shorting pins with the radial stub. Therefore, one will see different positioning of the bias inductors and pads across different designs. In this work, MACOM MAVR-000120-14110P (0.16pF - 1.1pF) are adopted as a tuning element.
%%%%%%%%%%%%%%%%%%%%%%%%%%%%%%%%%%%%%%%%%%%%%%%%%%%%%%%%%%%%%%%%%%%%%%%%%%%%%
\subsection{Design Examples}

In Fig. 2(b) and (c), tunable BPFs targeting 4G MB/HB are shown. In Fig 2(b), tuning range of 33\% (2.4 GHz - 2.8 GHz) is achieved with low insertion loss while maintaining return loss not exceeding -10 dB. It is observted that the notch left to the pass-band exploits the DC shorting effect thanks to the introduction of the additional shorting port. The varactor is able to tune the passband and the notch right to the passband. In Fig. 2(c), an alternative implementation of the layout is provided by replacing shorting pins with radial stubs. tuning range of 2.45 GHz to 2.95 GHz is observed with the insertion loss ranging from -1.2 dB to -1.3 dB. The fixed notch left to the pass-band is attributed to the  Wideband/narrow-band responses of the filters Fig. 2(b) are shown in Fig. 3. Low insertion loss from -1.5 dB to -2.2 dB is achieved and wide stopband with the rejection of level -20 dB ranges up to 10 GHz.

In Fig. 4, layout and simulated S-parameter responses of tunable BPFs targeting 5G n79 and UNII Bands are shown. It is noted that simulation results are obtained with the radial stub replacing the shorting pints associated with an ideal short load. The configuration is different from Fig. 2 in that now it does not incorporate none of the additional shorting ports as the open/open are chosen as the two randomized loads. Notably, the single varactor is able to tune both of the notches left/right to the pass band while the position of the passband is reconfigured properly. Tuning range of 4.64 GHz to 6.16 GHz (28\%) is achieved with insertion loss ranging from -2.6 dB to -3 dB is observed in Fig. 4(a) and the same tuning range with insertion loss ranging from -1.3 dB to -1.8 dB is observed in Fig. 4(b). The fractional bandwidth is higher in Fig. 4(b) than that of Fig. 4(a).

\section{Conclusion}
This work presented the novel 3-phase workflow to inversely design the pixelated bandpass filters with a single tuning varactor and a shorting load targeting 4G MB/HB to 5G n79 and UNII bands. The constructed bandpass filters show either exceptionally low loss compared to the state-of-the-art tunable counterparts, attributed to the single tuning element. Even though only the single tuning element is employed, tuning range is moderate thanks to the power of pixelated inverse design. To the authors' best of knowledge, this work marks the first demonstration of the microwave inverse design technique applied to the continuously tunable passive devices.

%%%%%%%%%%%%%%%%%%%%%%%%%%%%%%%%%%%%%%%%%%%%%%%%%%%%%%%%%%%%%%%%%%%%%%%%%%%%%

%%%%%%%%%%%%%%%%%%%%%%%%%%%%%%%%%%%%%%%%%%%%%%%%%%%%%%%%%%%%%%%%%%%%%%%%%%%%%

\bibliographystyle{IEEEtran}

\bibliography{IEEEabrv,IEEEexample}

@STRING{IEEE_J_MWCL       = "{IEEE} Microw. Wireless Compon. Lett."}

@STRING{IEEE_J_AP         = "{IEEE} Trans. Antennas Propag."}

@STRING{IEEE_J_MTT        = "{IEEE} Trans. Microw. Theory Techn."}

@STRING{IEEE_J_JSSC       = "{IEEE} J. Solid-State Circuits"}

@STRING{IEEE_M_MW         = "{IEEE} Microw. Mag."}

@article{two_1,
  title={Low-loss frequency-agile bandpass filters with controllable bandwidth and suppressed second harmonic},
  author={Zhang, Xiu Yin and Xue, Quan and Chan, Chi Hou and Hu, Bin-Jie},
  journal=IEEE_J_MTT,
  volume={58},
  number={6},
  pages={1557--1564},
  year={2010},
  publisher={IEEE}
}

@article{yang2015tunable,
  title={Tunable 1.25--2.1-GHz 4-pole bandpass filter with intrinsic transmission zero tuning},
  author={Yang, Tao and Rebeiz, Gabriel M},
  journal=IEEE_J_MTT,
  volume={63},
  number={5},
  pages={1569--1578},
  year={2015},
  publisher={IEEE}
}

@article{chen2018microstrip,
  title={Microstrip switchable and fully tunable bandpass filter with continuous frequency tuning range},
  author={Chen, Chi-Feng and Wang, Guo-Yun and Li, Jhong-Jhen},
  journal=IEEE_J_MWCL,
  volume={28},
  number={6},
  pages={500--502},
  year={2018},
  publisher={IEEE}
}

@article{hongkong,
  title={Compact Pixelated Printed Circuit Board-Based Bandpass Filter},
  author={Bi, Jingyun and Huang, Yixuan and Wang, Yanze and Zhao, Ruxuan and Zhou, Xinyu and Chan, Wing Shing},
  journal=IEEE_M_MW,
  year={2026},
  publisher={IEEE}
}

@article{woojun_arxiv,
  title={Inverse Design of Multi-Layered Manufacturable Pixelated Diplexers Through Optimized Geometrical Configuration and Meshing Strategy in MoM},
  author={Lee, Woojun and Lee, Jungmin and Hong, Ji Wu and Walling, Jeffrey S},
  journal=IEEE_J_MTT,
  year={2026},
  publisher={IEEE}
}

@inproceedings{jungmin_APMC,
  title={AI-Assisted EM Simulator For The Randomized Pixelated RF Components},
  author={Lee, Jungmin and Lee, Woojun and Hong, Ji Wu and Walling, Jeffrey S},
  booktitle={2025 Asia-Pacific Microwave Conference (APMC)},
  pages={1--3},
  year={2025},
  organization={IEEE}
}

@book{pozar2011microwave,
  title={Microwave engineering},
  author={Pozar, David M},
  year={2011},
  publisher={John wiley \& sons}
}

@article{port_reduction,
  title={Port reduction methods for scattering matrix measurement of an n-port network},
  author={Lu, Hsin-Chia and Chu, Tah-Hsiung},
  journal=IEEE_J_MTT,
  volume={48},
  number={6},
  pages={959--968},
  year={2000},
  publisher={IEEE}
}

@article{gamom,
  title={Genetic algorithms and method of moments (GA/MOM) for the design of integrated antennas},
  author={Johnson, J Michael and Rahmat-Samii, Yahya},
  journal=IEEE_J_AP,
  volume={47},
  number={10},
  pages={1606--1614},
  year={1999},
  publisher={IEEE}
}

@inproceedings{jungmin,
  title={Pixelated RF: Random Metasurface Based Electromagnetic Filters},
  author={Lee, Jungmin and Jia, Wei and Sensale-Rodriguez, Berardi and Walling, Jeffrey S},
  booktitle={2023 21st IEEE Interregional NEWCAS Conference (NEWCAS)},
  pages={1--5},
  year={2023},
  organization={IEEE}
}

@ARTICLE{prince,
  author={Karahan, Emir Ali and Liu, Zheng and Sengupta, Kaushik},
  journal=IEEE_J_JSSC, 
  title={Deep-Learning-Based Inverse-Designed Millimeter-Wave Passives and Power Amplifiers}, 
  year={2023},
  volume={58},
  number={11},
  pages={3074-3088},
  doi={10.1109/JSSC.2023.3276315}}

\end{document}